\documentclass[floats,onecolumn,superscriptaddress,floatfix]{revtex4}
\usepackage{epsf,graphicx}
\usepackage{amssymb}
\usepackage{amsmath}
\usepackage{eepic}
\usepackage[dvips]{color}
\usepackage{multirow}
\usepackage{color}
\begin{document}

\title{Heavy fermion behavior in the quasi-one-dimensional Kondo lattice CeCo$_2$Ga$_8$}

\author{Le Wang$^*$}
\affiliation{Beijing National Laboratory for Condensed Matter Physics, Institute of Physics, Chinese Academy of Sciences, Beijing 100190, China}
\affiliation{School of Physical Sciences, University of Chinese Academy of Sciences, Beijing 100190, China}
\author{Zhaoming Fu\footnote{These authors contributed equally to this work.}}
\affiliation{Beijing National Laboratory for Condensed Matter Physics, Institute of Physics, Chinese Academy of Sciences, Beijing 100190, China}
\affiliation{College of Physics and Material Science, Henan Normal University, Xinxiang 453007, China}
\author{Jianping Sun}
\affiliation{Beijing National Laboratory for Condensed Matter Physics, Institute of Physics, Chinese Academy of Sciences, Beijing 100190, China}
\affiliation{School of Physical Sciences, University of Chinese Academy of Sciences, Beijing 100190, China}
\author{Min Liu}
\author{Wei Yi}
\affiliation{Beijing National Laboratory for Condensed Matter Physics, Institute of Physics, Chinese Academy of Sciences, Beijing 100190, China}
\author{Changjiang Yi}
\affiliation{Beijing National Laboratory for Condensed Matter Physics, Institute of Physics, Chinese Academy of Sciences, Beijing 100190, China}
\affiliation{School of Physical Sciences, University of Chinese Academy of Sciences, Beijing 100190, China}
\author{Yongkang Luo}
\author{Yaomin Dai}
\affiliation{Los Alamos National Laboratory, Los Alamos, New Mexico 87545, USA}
\author{Guangtong Liu}
\affiliation{Beijing National Laboratory for Condensed Matter Physics, Institute of Physics, Chinese Academy of Sciences, Beijing 100190, China}
\author{Yoshitaka Matsushita}
\affiliation{Materials Analysis Station, National Institute for Materials Science, 1-2-1 Sengen, Tsukuba, Ibaraki 305-0047, Japan.}
\author{Kazunari Yamaura}
\affiliation{Research Center for Functional Materials, National Institute for Materials Science, 1-1 Namiki, Tsukuba, Ibaraki 305-0044, Japan}
\author{Li Lu}
\affiliation{Beijing National Laboratory for Condensed Matter Physics, Institute of Physics, Chinese Academy of Sciences, Beijing 100190, China}
\affiliation{School of Physical Sciences, University of Chinese Academy of Sciences, Beijing 100190, China}
\affiliation{Collaborative Innovation Center of Quantum Matter, Beijing 100190, China}
\author{Jin-guang Cheng}
\affiliation{Beijing National Laboratory for Condensed Matter Physics, Institute of Physics, Chinese Academy of Sciences, Beijing 100190, China}
\author{Yi-feng Yang}
\email[]{yifeng@iphy.ac.cn}
\affiliation{Beijing National Laboratory for Condensed Matter Physics, Institute of Physics, Chinese Academy of Sciences, Beijing 100190, China}
\affiliation{School of Physical Sciences, University of Chinese Academy of Sciences, Beijing 100190, China}
\affiliation{Collaborative Innovation Center of Quantum Matter, Beijing 100190, China}
\author{Youguo Shi}
\email[]{ygshi@iphy.ac.cn}
\affiliation{Beijing National Laboratory for Condensed Matter Physics, Institute of Physics, Chinese Academy of Sciences, Beijing 100190, China}
\affiliation{School of Physical Sciences, University of Chinese Academy of Sciences, Beijing 100190, China}
\author{Jianlin Luo}
\affiliation{Beijing National Laboratory for Condensed Matter Physics, Institute of Physics, Chinese Academy of Sciences, Beijing 100190, China}
\affiliation{School of Physical Sciences, University of Chinese Academy of Sciences, Beijing 100190, China}
\affiliation{Collaborative Innovation Center of Quantum Matter, Beijing 100190, China}

\maketitle

\section*{ABSTRACT}
Dimensionality plays an essential role in determining the anomalous non-Fermi liquid properties in heavy fermion systems. So far most heavy fermion compounds are quasi-two-dimensional or three-dimensional. Here we report the synthesis and systematic investigations of the single crystals of the quasi-one-dimensional Kondo lattice CeCo$_2$Ga$_8$. Resistivity measurements at ambient pressure reveal the onset of coherence at $T^*\approx 20\,$K and non-Fermi liquid behavior with linear temperature dependence over a decade in temperature from 2 K to 0.1 K. The specific heat increases logarithmically with lowering temperature between 10 K and 2 K and reaches 800 mJ/mol K$^2$ at 1 K, suggesting that CeCo$_2$Ga$_8$ is a heavy fermion compound in the close vicinity of a quantum critical point. Resistivity measurements under pressure further confirm the non-Fermi liquid behavior in a large temperature-pressure range. The magnetic susceptibility is found to follow the typical behavior for a one-dimensional (1D) spin chain from 300 K down to $T^*$, and first-principles calculations predict flat Fermi surfaces for the itinerant $f$-electron bands. These suggest that CeCo$_2$Ga$_8$ is a rare example of the quasi-1D Kondo lattice, but its non-Fermi liquid behaviors resemble those of the quasi-two-dimensional YbRh$_2$Si$_2$ family. The study of the quasi-one-dimensional CeCo$_2$Ga$_8$ family may therefore help us to understand the role of dimensionality on heavy fermion physics and quantum criticality.

\section*{INTRODUCTION}
Heavy fermion materials exhibit a rich variety of exotic correlated states, such as non-Fermi liquid and unconventional superconductivity, which often occur in the vicinity of a magnetic quantum critical point \cite{White2015,Steglich2016,Yang2016}. The special non-Fermi liquid behavior may be compared to experiment to reveal the effective low-energy physics of the correlated electrons \cite{Coleman2005,Si2010}. Taking the 2D antiferromagnetism as an example, the conventional spin density wave (SDW) scenario based on the fluctuations of the magnetic order parameter has predicted as a function of temperature the resistivity, $\rho\sim T$, and the specific heat, $C/T\sim -\ln T$, in the quantum critical regime \cite{Lohneysen2007}. While these have been confirmed in some heavy fermion compounds such as CeCu$_2$Si$_2$, disparate temperature dependences of the resistivity and the specific heat have been observed in both stoichiometric and slightly Ge-doped YbRh$_2$Si$_2$ as the systems approach the quantum critical point \cite{Stockert2011}. This indicates the failure of the conventional SDW scenario in describing the low-energy physics and points to possible breakup of the composite (heavy) quasiparticles near the antiferromagnetic quantum critical point in the YbRh$_2$Si$_2$ family. 

Such unconventional quantum criticality in heavy fermion systems defy a satisfactory theoretical understanding because of the non-perturbative nature of the Kondo lattice physics despite of many experimental and theoretical efforts in recent years. The well-known mean-field approximation only captures the hybridization physics qualitatively, but fails to provide more insight into the detailed nature of the quantum criticality \cite{Coleman2007}. Exact numerical methods such as the density matrix renormalization group approach \cite{Schollwock2005} are typically limited to one-dimensional (1D) systems, while most heavy fermion materials exhibit strong quasi-two-dimensional (2D) or three-dimensional (3D) features \cite{Stewart2001,Krellner2011,Steppke2013}. Dimensionality is a key ingredient in determining the non-Fermi liquid behavior. The study of quasi-1D systems may be useful for our understanding of the heavy fermion physics and the associated unconventional quantum critical behavior. 

In this work, we report experimental studies on the single crystal CeCo$_2$Ga$_8$ and show that it is a quasi-1D Kondo lattice system with a quantum critical point near ambient pressure. After the successful synthesis of the single crystals, we noticed that a polycrystalline CeCo$_2$Ga$_8$ has been reported previously \cite{Koterlin1989}, in which electrical resistivity and thermoelectric power have been measured but shown no interesting properties. In this work we carried out more detailed investigations of the single crystal CeCo$_2$Ga$_8$. A picture of the single crystal is shown in Figure 1(a). The compound adopts the YbCo$_2$Al$_8$-type orthorhombic structure (space group Pbam, No. 55) with the lattice parameters $a$ = 12.3792(7) \r{A}, $b$ = 14.3053(9) \r{A}, and $c$ = 4.0492(3) \r{A}. The whole structure can be viewed as built of fused polyhedral of Ga that are interstitially stabilized by Co and Ce. The Co atoms are situated in tricapped trigonal prisms formed by nine Ga atoms. The Ce atoms form a chain along the $c$-axis located in the center of the pentagon formed by five CoGa$_9$ cages in the $ab$ plane. The Ce-Ce distances between neighboring chains are about 6.5 \r{A} and 7.5 \r{A}, much longer than the intra-chain distance of 4.05 \r{A}, indicating that it might be a quasi-1D system. As a result, the single crystal prefers to grow up along the $c$-axis as shown in Fig. 1. 

\begin{figure}[t]
\includegraphics[width=0.5\textwidth]{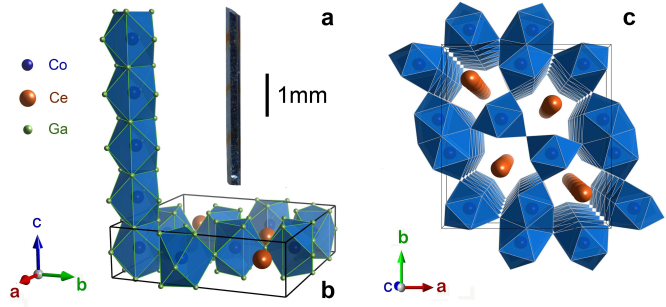}
\caption{\label{fig1}(Color online) \textbf{Crystalline structure of CeCo$_2$Ga$_8$.} (\textbf{a}) Picture of a single crystal of CeCo$_2$Ga$_8$ of about 4 mm in length along the $c$-axis. (\textbf{b}) Side view of the YbCo$_2$Al$_8$-type orthorhombic structure with face-shared CoGa$_9$ cages along the $c$-axis. (\textbf{c}) Stereoscopic view of the crystal structure showing the quasi-1D chains of Ce atoms along the $c$-axis. Each unit cell contains four Ce atoms and the interchain Ce-Ce distances are 6.5 \r{A} and 7.5 \r{A}, respectively.}
\end{figure}

Our measurements reveal characteristic non-Fermi liquid behavior in the normal state where the resistivity exhibits linear temperature dependence between 0.1 K and 2 K and the specific heat grows logarithmically with lowering temperature between 2 K and 10 K. These non-Fermi liquid behaviors are similar to those observed in YbRh$_2$Si$_2$ despite of the different dimensionality of the two compounds, i.e. quasi-1D vs quasi-2D, and may therefore be of potential interest for future investigations. The resistivity shows a coherence peak at $T^*\approx 20\,$K. Above $T^*$, the Hall coefficient obeys the skew scattering theory as seen in most heavy fermion compounds \cite{Fert1987}. The magnetic susceptibility deviates from the usual Curie-Weiss law below about 150 K but could be well fitted with the typical 1D Bonner-Fisher formula \cite{Bonner1964} from $300\,$K down to $T^*$, implying the quasi-one dimensionality of the underlying Kondo lattice. Our first-principles calculations also yield several flat Fermi sheets originating from the quasi-1D $f$-electron bands along the $c$-axis. Combining these experimental and theoretical results, it suggests the quasi-1D nature of the heavy fermion compound CeCo$_2$Ga$_8$, which provides an interesting basis for future investigation of the Kondo lattice physics in 1D. Resistivity measurements under pressure confirm the quantum critical behavior in a large temperature-pressure range.

\begin{figure}[t]
\includegraphics[width=0.45\textwidth]{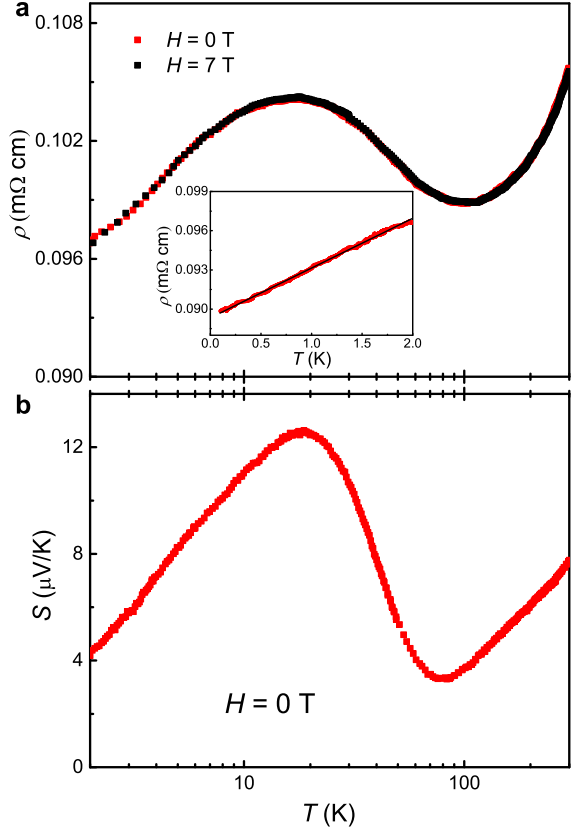}
\caption{\label{fig2}(Color online) \textbf{Temperature dependence of the electrical resistivity and Seebeck coefficient.} (\textbf{a}) Electrical resistivity $\rho$ of CeCo$_2$Ga$_8$ single crystal along the $c$-axis at zero field and $H=7\,$T. The two sets of data are almost superimposable, showing small magnetoresistance. The inset plots the resistivity below 2 K measured in a dilution refrigerator, indicating linear-in-$T$ (non-Fermi-liquid) behavior over a decade in temperature from 2 K down to 0.1 K. No sign of superconductivity is observed down to 0.1 K. (\textbf{b}) Temperature dependence of the Seebeck coefficient at zero field, showing similar behavior as in the resistivity.}
\end{figure}

\section*{RESULTS}
\noindent\textbf{Resistivity and Seebeck coefficient}\\
Figure 2 presents the temperature-dependence of the resistivity $\rho$ and the Seebeck coefficient $S$ along the $c$-axis. Both curves show similar logarithmic temperature dependence between about 20 K and 90 K originating from the incoherent Kondo scattering \cite{Kondo1964} of the conduction electrons by the localized $f$-moments. A broad peak appears at $T^*\approx 20$ K, which marks a crossover from the insulating-like behavior to the metallic behavior at lower temperatures. Above 90 K, the resistivity also exhibits metallic behavior where the Kondo scattering is suppressed and the transport property is governed by the electron-phonon scattering. These features are common in Ce-based heavy fermion materials. Importantly, as shown in the inset of Fig. 2(a), the resistivity exhibits $T$-linear behavior over a decade in temperature from 2 K to 0.1 K, in resemblance of those found in CeCoIn$_5$ and YbRh$_2$Si$_2$ \cite{Sidorov2002,Steglich2014}. This indicates that the single crystal CeCo$_2$Ga$_8$ locates near a quantum critical point \cite{Coleman2005,Si2010,Stockert2011}. However, we observe no sign of superconductivity down to 0.1 K. 

We note that our sample exhibits a large residual resistivity, $\rho_0\approx 90\,\mu\Omega\,$cm, and a small residual resistivity ratio, $\text{RRR}\approx 1.2$. To improve the quality of the single crystals, we have grown many samples but found similar value for the RRR while $\rho_0$ could be reduced by a factor of 2 (see, for example, Figure 3(a), where $\rho_0\sim 60\,\mu\Omega\,$cm and RRR $\sim1.1$ at ambient pressure). This seems to suggest that the small RRR might be an intrinsic property of this compound at ambient pressure. Closeness to a quantum critical point and small values of $T^*$ may lead to a large $\rho_0$. However, this contribution typically decreases rapidly with pressure, in contrast to the weak pressure dependence of our measured $\rho_0$. We speculate that the large $\rho_0$ might be partly due to the impurity (Ga) scattering enhanced by the quasi-1D nature of the charge transport in CeCo$_2$Ga$_8$. However, our single crystal X-ray diffraction data show no significant site/chemical disorder. More work are needed to further improve the sample quality and solve this issue.

\begin{figure}[t]
\includegraphics[width=0.45\textwidth]{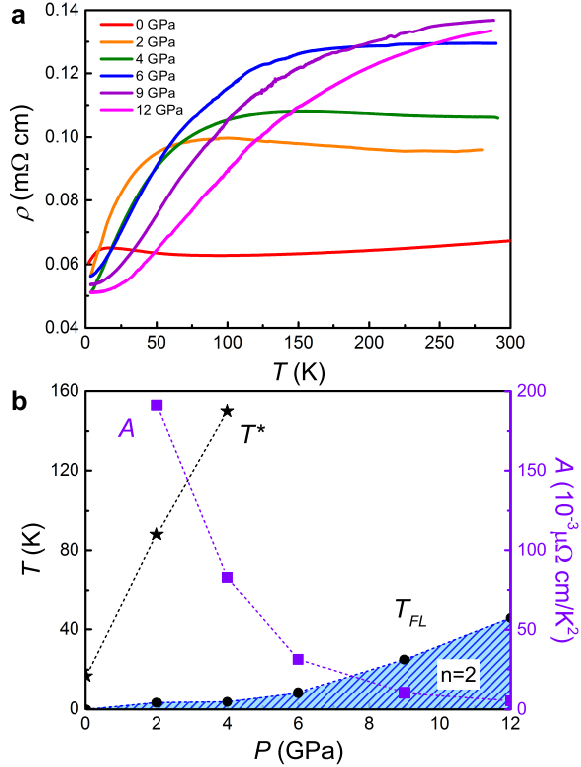}
\caption{\label{fig3}(Color online) \textbf{Pressure dependence of the $c$-axis resistivity as a function of temperature.} (\textbf{a}) Temperature dependence of the resistivity under pressure. (\textbf{b}) Temperature-pressure phase diagram derived from the resistivity data, where the coherence temperature, $T^*$, and the Fermi liquid temperature, $T_{FL}$, are both plotted for comparison. Details on the determination of the Fermi liquid regime ($n=2$) can be found in the Supplementary Fig.~S1. Also shown is the pressure dependence of the resistivity coefficient, $A$, defined as $\rho-\rho_0=A T^2$ in the Fermi liquid regime.
}
\end{figure}

The possible existence of impurities raises the question concerning the origin of the power-law behavior in the resistivity. To explore this, we further performed pressure measurements of the resistivity (on a different sample). Fig. 3(a) plots the pressure dependence of the $c$-axis resistivity as a function of temperature. As shown in the Supplementary Figure S1, our detailed analysis on the power-law behavior of the resistivity reveals continuous variation of the upper and lower boundaries of the non-Fermi liquid regime ($n=1$, where $n$ is the resistivity exponent defined as, $\rho \sim T^n$). The large temperature range of the non-Fermi liquid regime is not unusual. In CeCoIn$_5$, the upper boundary goes above 10 K with $T^*\sim 50\,$K at ambient pressure \cite{Sidorov2002}, while in CeRhIn$_5$, it follows roughly $T^*/2$ at high pressures \cite{Park2011}. The pressure variation of the coherence temperature, $T^*$, estimated from the resistivity peak and the Fermi liquid temperature, $T_{FL}$, obtained from the upper boundary of the Fermi liquid regime ($n=2$), are plotted in Fig. 3(b). We see that both $T^*$ and $T_{FL}$ increase with increasing pressure, consistent with the usual expectation for the Ce-based heavy fermion compounds. The fact that $T_{FL}$ approaches zero below 2$\,$GPa indicates that CeCo$_2$Ga$_8$ at ambient pressure indeed locates near a quantum critical point. In particular, the resistivity coefficient, $A$, defined as $\rho-\rho_0=A T^2$ in the Fermi liquid regime, increases by almost two orders of magnitude from 12 GPa to 2 GPa and tends to diverge approaching the ambient pressure. Since the effective mass of heavy quasiparticles follows $m^* \propto A^{1/2}$. This indicates that the heavy electrons also have an enhanced effective mass, as has been observed in CeRhIn$_5$ near the pressure-induced quantum critical point. These results are typical for Ce-based heavy fermion compounds.\\

\begin{figure}[t]
\includegraphics[width=0.45\textwidth]{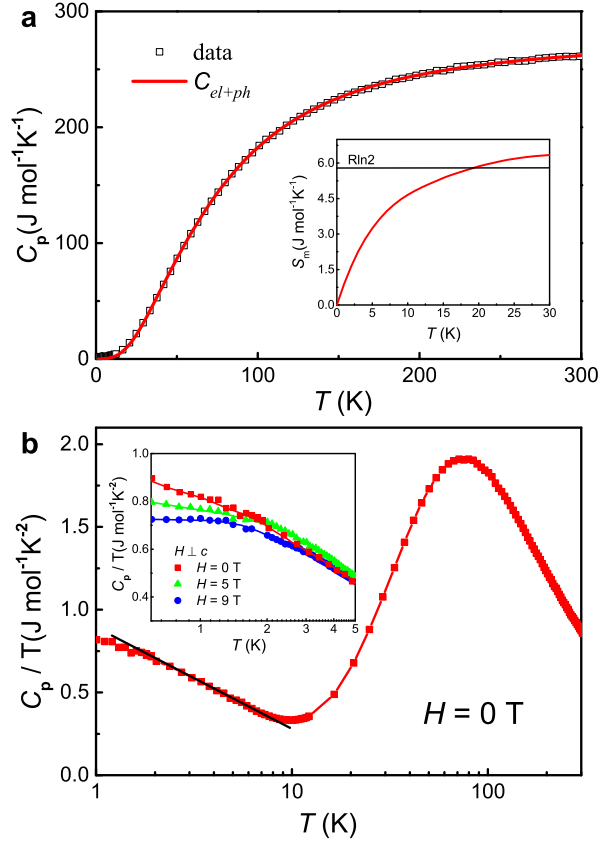}
\caption{\label{fig4}(Color online) \textbf{Temperature dependence of the specific heat.} (\textbf{a}) The zero-field specific heat $C_p$ as a function of temperature on a linear scale. The solid line represents the theoretical fit between 30 K and 300 K using a Debye-Einstein model for the phonon contribution. The inset shows the magnetic entropy calculated from the specific heat data after subtracting the phonon contribution. (\textbf{b}) Temperature dependence of $C_p/T$ on a semilogarithmic scale, showing a logarithmic divergence between 2 K and 10 K. The inset plots the field variation of $C_p/T$ at low temperatures for $H=0,\,5,\,9\,$T in perpendicular to the $c$-axis.}
\end{figure}

\noindent\textbf{Heat capacity}\\
Figure 4(a) plots the measured specific heat $C_p$ as a function of temperature at zero field. At $T=300$ K, $C_p$ is about 260.8 J/mol K, close to the Dulong-Petit limit \cite{Kittel1966}, $3nR=274.2$ J/mol K, where $n=11$ is the total number of atoms per formula unit and $R$ is the gas constant. The high temperature specific heat above 30 K can be well fitted taking into account the contributions of conduction electrons and phonons \cite{Yang2015,Prakash2016}, which yields a specific heat coefficient of the conduction electrons, $\gamma_0=6.3\,$mJ/mol K$^2$. We note that this fit is just a theoretical approximation. The background electronic and phonon contributions should be better compared with that of a nonmagnetic reference compound. Unfortunately, we have failed to grow the single crystal LaCo$_2$Ga$_8$. Yet our first-principles calculations for the hypothetical LaCo$_2$Ga$_8$ crystal with the same lattice structure yield $\gamma_0^{th}=8.4\,$mJ/mol K$^2$, in rough agreement with the above fitting result. The inset of Fig. 4(a) shows the magnetic entropy $S_m$ after subtracting the above electronic and phonon contributions. We see $S_m$ becomes saturated and reaches $R\ln2$ at about 20 K, close to $T^*$ determined from the resistivity peak \cite{Yang2008}. This implies a doublet ground state for the Ce $f$-electrons. Fig. 4(b) plots $C_p/T$ as a function of the temperature on a semilogarithmic scale. We see a minimum at about 10 K that separates clearly the low temperature behavior from the high temperature phonon contributions. Below 10 K, $C_p/T$ grows logarithmically with lowering temperature down to 2 K, a characteristic signature of quantum criticality. An extrapolation of the logarithmic behavior to high temperatures also gives an onset temperature of about 20 K, consistent with the previously determined $T^*$ from the resistivity peak. This indicates that $T^*$ indeed marks the onset of lattice coherence. At 1 K, the specific heat coefficient reaches $\sim$ 800 mJ/mol K$^2$, a hundred times of that of the background conduction electrons but comparable to those of typical heavy fermion compounds such as CeCoIn$_5$ \cite{Petrovic2001}. Applying a magnetic field of 9 T suppresses the divergence at low temperatures and gives rise to a Fermi liquid state with roughly temperature-independent $C_p/T$, as shown in the inset of Fig. 4(b).\\

\begin{figure}
\includegraphics[width=0.45\textwidth]{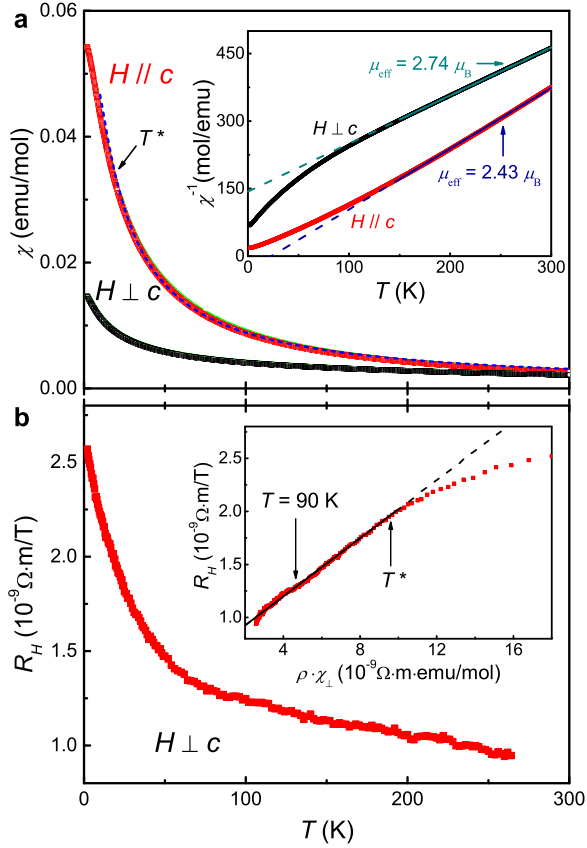}
\caption{\label{fig5}(Color online) \textbf{Temperature dependence of the susceptibility and Hall coefficient.} (\textbf{a}) The FC (red and black) and ZFC (green) susceptibility of CeCo$_2$Ga$_8$ along ($\chi_\parallel$) and perpendicular ($\chi_\perp$) to the $c$-axis. The dashed line is the fit to $\chi_\parallel$ using the formula of 1D spin chain. The inset shows the the Curie-Weiss fit (dashed lines) to the inverse susceptibility, yielding $\mu_{eff}=2.74\ \mu_B$ for $H\perp{c}$ and $\mu_{eff}=2.43\ \mu_B$ for $H\parallel{c}$ at high temperatures. (\textbf{b}) The Hall coefficient ($H\perp{c}$, $j \parallel c$) as a function of temperature. The inset compares the Hall coefficient with the prediction (dashed line) of the skew scattering theory.}
\end{figure}

\noindent\textbf{Magnetic susceptibility and Hall coefficient}\\ 
Figure 5 gives the measured magnetic susceptibility and the Hall coefficient. The field-cooling (FC) and zero-field-cooling (ZFC) data are essentially superimposable. The susceptibilities increase monotonically with decreasing temperature, showing no sign of magnetic transitions. However, the magnetization for $H\parallel{c}$ is nearly 3 times as large as that of $H\perp{c}$ at $T=2$ K, as shown in the Fig. 5(a). This indicates a strong anisotropy in CeCo$_2$Ga$_8$ and the $c$-axis is the easy axis. Above 150 K, a Curie-Weiss fit using $\chi(T)=C/(T-\theta)$, where $C$ is the Curie constant and $\theta$ is the Weiss temperature, yields the moment $\mu_{eff}=2.74\ \mu_B$ for $H\perp{c}$ and $\mu_{eff}=2.43\ \mu_B$ for $H\parallel{c}$, close to the free-ion moment of Ce$^{3+}$, 2.54 $\mu_B$. However, the fit fails below $T=150$ K, well above the coherence temperature, $T^*\approx20$ K. While this has been observed in many heavy fermion materials and often ascribed to the crystal field effect, we find here that an alternative formula for 1D spin chain \cite{Bonner1964}, $\chi(T)=C/T e^{-\eta|J|/T}$ where $J$ is the exchange coupling and $\eta$ accounts for the anisotropy, could yield an excellent fit for $\chi_\parallel$ from 300 K down to $T^*$, giving an effective antiferromagnetic coupling $\eta|J|=6.9\,$ K and an effective moment $\mu_{eff}=2.74\ \mu_B$, in agreement with the Curie-Weiss fit at high temperatures. This suggests that the $f$-electrons in CeCo$_2$Ga$_8$ are well localized and form a quasi-1D spin chain at high temperatures, consistent with the quasi-1D crystal structure shown in Fig. 1. The $c$-axis Hall coefficient is also measured and presented in Fig. 5(b). We find $R_H$ is proportional to $\rho\chi_\perp$ above $T^*$ and governed by the incoherent skew scattering following the typical behavior in most other heavy fermion systems \cite{Fert1987}. The deviation below $T^*$ signals the onset of the $f$-electron coherence.\\

\begin{figure}[t]
\includegraphics[width=0.5\textwidth]{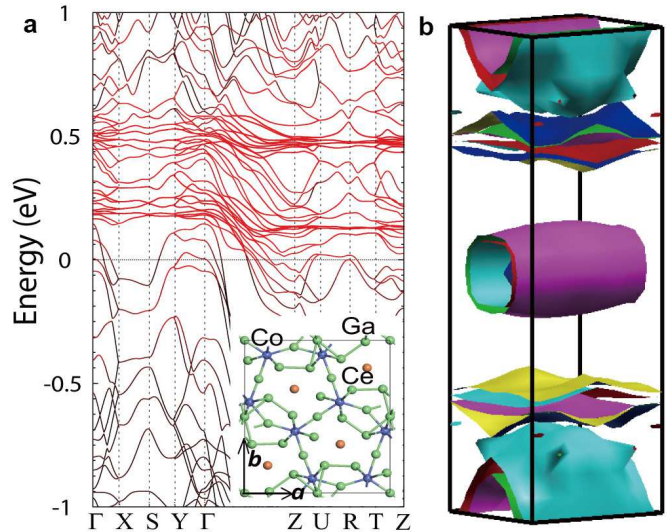}
\caption{\label{fig6}(Color online) \textbf{First-principles calculations for the band structures and Fermi surfaces.} (\textbf{a}) The band structures, showing flat $f$-electron bands within the $ab$ plane and a spin-orbit splitting of about 0.3 eV. The inset shows the unit cell containing 4 formula units and a total number of 44 atoms. (\textbf{b}) The Fermi surfaces showing four flat sheets from the four Ce $f$-chains along the $c$-axis and several cylindrical sheets from the Co/Ga layered structure perpendicular to the $b$-axis. The colors have no special meaning and are generated automatically from the WIEN2K code to distinguish different Fermi sheets.}
\end{figure}

\noindent\textbf{First-principles calculations}\\ 
The quasi-1D nature of the underlying Kondo lattice is further supported by the first-principles density functional theory calculations (DFT) using the full-potential linearized augmented-plane-wave method as implemented in the WIEN2K package \cite{Blaha2001}. Although DFT cannot treat correctly the strong electronic correlations, it usually yields qualitatively good predictions on the Fermi surface topology of the  $f$-electron bands and has therefore been widely used as a starting point for understanding the electronic properties of many heavy fermion compounds. Calculations using strongly correlated methods such as the dynamical mean-field theory are too time-consuming for this compound as its unit cell contains 4 formula units with a total number of 44 atoms including 4 Ce-ions, as shown in the inset of Figure 6(a). Our DFT calculations treat the $f$-electrons as fully itinerant and take into account the spin-orbit coupling. We use the refined lattice parameters as listed in the Supplementary Table S1 and S2. The GGA-PBE functional \cite{Perdew1996} is adopted for the exchange correlation energy with $R_{MT}\times K_{max}$ = 8.0 and 1000 $k$-point meshes over the Brillouin zone. The obtained band structures and Fermi surfaces are presented in Fig. 6. We see that the Ce $f$-bands split into $J = 5/2$ and $J=7/2$ multiplets with an energy difference of about 0.3 eV. The $J= 5/2$ multiplet have the lower energy and locate slightly above the Fermi energy. Importantly, we see that the itinerant $f$-bands are only dispersive along the $\Gamma-Z$ path in the Brillouin zone (namely the $c$-axis in real space), but remain flat in the $ab$ plane, indicating the quasi-1D property of the itinerant $f$-electrons. The Fermi surfaces show two different types of topologies, including four flat sheets originating from the quasi-1D $f$-electron bands of the four Ce ions in each unit cell and quasi-two dimensional cylindrical Fermi surfaces from the conduction electrons owing to the layered structure perpendicular to the $b$-axis. As may have been clearly seen in Fig. 1, the heavy electron physics originates from the hybridization between the Ce $f$-chains and the conduction electrons from the surrounding CoGa$_9$-cages. Our calculations yield a specific heat coefficient of about 22.9 mJ/mol K$^2$, much smaller than the experimentally observed value of $\sim$ 800 mJ/mol K$^2$, indicating the great enhancement due to quantum criticality. 

\section*{DISCUSSION AND CONCLUSION}
We have synthesized the single crystals of CeCo$_2$Ga$_8$ and carried out systematic investigations of its magnetic, thermodynamic and transport properties. We find that CeCo$_2$Ga$_8$ behaves like a quasi-1D Kondo lattice and locates in the close vicinity of a magnetic quantum critical point. It exhibits non-Fermi liquid behavior with a $T$-linear resistivity below 2 K and a logarithmically divergent specific heat between 2 K and 10 K at ambient pressure. In the conventional Hertz-Millis theory of quantum criticality, these scaling behaviors are predicted for the 2D antiferromagnetic quantum critical point but not expected in a quasi-1D Kondo lattice system. It will be crucial to examine the dimensionality of the critical spin fluctuations. On the other hand, the observed disparate temperature dependences in the two quantities below 2 K are very similar to experimental observations in YbRh$_2$Si$_2$ \cite{Steglich2014}. These suggest a possibly unconventional quantum critical point whose nature is yet to be explored \cite{Stockert2011}. We note that there are also other Ce-128 compounds including CeCo$_2$Al$_8$, CeFe$_2$Ga$_8$, and CeFe$_2$Al$_8$ \cite{Koterlin1989,Koterlin1994,Kolenda2001,Ghosh2012}, none of which has been well studied. Synthesis of good single crystals seems more challenging in this family of Ce-128 compounds compared to the famous Ce-115 family, possibly due to their structural differences. Nevertheless, a systematic exploration of the whole family is worthwhile for future investigations. The effect of chemical tuning may help us to achieve a better understanding of the Kondo lattice physics and, in particular, the role of dimensionality on heavy fermion quantum criticality.

\section*{METHODS}
\noindent\textbf{Sample preparation and characterization}\\
Single crystals of CeCo$_2$Ga$_8$ were grown using a Ga self-flux method in alumina crucible which was sealed in a fully evacuated quartz tube. The crucible was heated to 1100 $^\circ$C for 10 h, then cooled slowly to 630 $^\circ$C at which point the Ga flux was spun off in a centrifuge, and subsequently quenched in cold water. Rodlike single crystals were yielded with the length of $\sim$ 4 mm. Elemental analysis was conducted via energy dispersive X-ray (EDX) spectroscopy using a Hitachi S-4800 scanning electron microscope at an accelerating voltage of 15 kV, with an accumulation time of 90 s. The result of EDX indicated the composition of CeCo$_2$Ga$_8$ was stoichiometric. Single crystal X-ray diffraction was carried out on a RIGAKU Saturn CCD Diffractometer with a VariMax confocal optical system at 213(2) K using Mo $K_{\alpha}$ radiation ($\lambda$ = 0.71073 \r{A}). The crystal structure was refined by full-matrix least-squares fitting on {\sl F$^2$} using the SHELXL-2014/7 program.\\ 

\noindent\textbf{Transport, heat capacity and magnetic measurements}\\
The magnetic susceptibility ($\chi$) was measured in a Quantum Design Magnetic Property Measurement System (MPMS) from 2 K to 300 K under various applied magnetic fields up to 50 kOe in field-cooling (FC) and zero-field-cooling (ZFC) modes. A well crystallized sample was picked out for the study of magnetic anisotropy with the field perpendicular to and along the $c$-axis, respectively. The specific heat was measured in a Physical Property Measurement System (PPMS) with He-3 option. The electrical resistivity ($\rho$) along the $c$-axis was measured in PPMS  upon cooling from 300 K to 2 K and in a top-loading dilution refrigerator using the standard low frequency lock-in technique below 2 K. High-pressure resistivity measurements up to 12 GPa were performed with a standard four-probe method in a palm cubic anvil cell apparatus using glycerol as pressure transmitting medium (PTM) in order to maintain a good hydrostatic pressure condition \cite{Cheng2014}. 

\section*{ACKNOWLEDGEMENTS}
This work was supported by the National Natural Science Foundation of China (Grant Nos. 11274367, 11522435, 11474330, 11574377), the National Key Research and Development Program of China (2016YFA0300604), the State Key Development Program for Basic Research of China (2015CB921300, 2014CB921500), and the Strategic Priority Research Program (B) of the Chinese Academy of Sciences (Grant Nos. XDB07020000, XDB07020200, XDB07020100). Work in Japan was supported by the Japan Society for the Promotion of Science (JSPS) through a Grant-in-Aid for Scientific Research (15K14133, 16H04501). Work at Los Alamos was performed under the auspices of the US Department of Energy, Division of Materials Sciences and Engineering.\\

\section*{COMPETING INTERESTS}
The authors declare no competing financial interests.\\

\section*{CONTRIBUTIONS}
Y.Y. and Y.S. conceived the idea and supervised the project; L.W. and Y.S. synthesized the single crystals; L.W., J.S., W.Y., C.Y., Y.L., Y.D., G.L., Y.M., K.Y., L.L., J.C., Y.S. and J.L. performed the measurements; Z.F., M.L. and Y.Y. did the theoretical analysis; all authors discussed the results; L.W., Z.F., Y.Y. and Y.S. wrote the paper.\\

\renewcommand{\thefigure}{S\arabic{figure}}
\renewcommand\thetable{S\arabic{table}}
\setcounter{figure}{0} 
\setcounter{table}{0} 

\begin{table}[h]
\caption{\label{Supplementary table S1}Crystallographic data and structure refinement for CeCo$_2$Ga$_8$}
	\begin{tabular}{ll}		
		\hline
		empirical formula & CeCo$_2$Ga$_8$     \\
		formula weight & 815.74 g/mol     \\
		temperature & 213(2) K  \\
		wavelength  & Mo $K_\alpha$ (0.71073 \AA{}) \\
		crystal system & orthorhombic \\
		space group & Pbam (55) \\
		unit cell dimensions & $a=12.3792(7)$\AA{} \\
		& $b=14.3053(9)$\AA{} \\
		& $c=4.0492(3)$\AA{} \\ 
		cell volume	& 717.07(8) \AA{}$^3$ \\
		$Z$ & 4 \\
		density, calculated & 7.55573 g/cm$^3$ \\
		crystal size (mm)	& $0.099 \times 0.050 \times 0.035$ \\
		$h \ k \ l$ range & $-25 \le h \le 25, -20 \le k \le 29, -6 \le l \le 8$ \\
		2$\theta_{max}$ & 92.54 \\
		linear absorption coeff. & 40.119 mm$^{-1}$ \\
		absorption correction & multi-scan \\
		$T_{min}/T_{max}$ & 0.1200/0.2894 \\
		no. of reflections & 18374 \\
		$R_{int}$ & 0.0548 \\
		no. independent reflections & 3398 \\
		no. observed reflections & 3021[$F_o > 4\sigma (F_o)$] \\
		$F$(000) & 1440 \\
		$R$ values$^\alpha$ & 4.91 \% ($R_1[F_o > 4\sigma (F_o)]$)		\ \ \ 12.02 \% (w$R_2$) \\
		weighting scheme &　$w=1/[\sigma^2(F_o^2) + (0.0676P)^2 + 1.6858P]$, where $P = (F_o^2 + 2F_c^2)/3$　\\
		diff. Fourier residues & [-3.474,14.794] e/\AA{}$^3$ \\
		refinement software & SHELXL-2014/7 \\	
		\hline		
	\end{tabular}

	 \begin{flushleft}
	  \ \ \ \ \ \ \ \ \ \ \ \ \ \ \ \ \ \ \ \ \ $^\alpha$More than double difference in the $R$ values are probably due to the weighting scheme used in the analysis.	
	 \end{flushleft}
\end{table} 

\newpage 
\begin{table}[h]
	\caption{\label{table2}Structure parameters and anisotropic displacement parameters (\AA{}$^2$) of CeCo$_2$Ga$_8$}
	\begin{tabular}{lcllll}
	\hline	
	Site \ \ \ \ \ \  & Wyckoff position \ \ \ \ \ \  & x \ \ \ \ \ \ \ \ \ \ \ \ \ \ \ \ \ \ \ \ \ & y  \ \ \ \ \ \ \ \ \ \ \ \ \ \ \ \ \ \ \ \ \ & z \ \ \ \ \ \ \ \ \ \ \ \ \ \ \	& $U_{eq}$ \\
	\hline
	Ce & 4h & 0.34193(2) & 0.68065(2) & 0.5000 & 0.00700(6) \\
    Co1	& 4h &　0.53441(4)　& 0.90574(4) & 0.5000 & 0.00532(9)	\\
    Co2	&4h	& 0.65323(4)	& 0.59633(4) &	0.5000	& 0.00449(9) \\
    Ga1	&2c	&0.5000	&1.0000	&0.0000	&0.00677(10) \\
    Ga2	&4g	&0.45038(4)	&0.81869(3)	&0.0000	&0.00660(8) \\
    Ga3	&4g	&0.66282(3)	&0.87809(4)	&0.0000	&0.00640(8) \\
    Ga4	&4h	&0.59690(3)	&0.75312(3)	&0.5000	&0.00727(8) \\
    Ga5	&4g	&0.52270(4)	&0.63216(3)	&0.0000	&0.00616(8) \\
    Ga6	&4g	&0.73769(4)	&0.67549(3)	&0.0000	&0.00682(8) \\
    Ga7	&2b	&0.5000	&0.5000	&0.5000	&0.00804(11) \\
    Ga8	&4g	&0.67005(4)	&0.49036(4)	&0.0000	&0.00704(8) \\
    Ga9	&4h	&0.83926(3)	&0.54461(4)	&0.5000	&0.00821(9) \\
    \hline		
	\end{tabular}
	\begin{tabular}{lllllll}
	Atom \ \ \ \ \ \ \ 	&U11 \ \ \ \ \ \ \ \ \ \ \ \ \ &U22 \ \ \ \ \ \ \ \ \ \ \ \ \ &U33 \ \ \ \ \ \ \ \ \ \ \ \ \ 	&U12 \ \ \ \ \ \ \ \ \ \ \ \ \ &U13 \ \ \ \ \ \ \ \ \ \ \ \ &U23 \\
	\hline	
	Ce	&0.00621(9)	&0.00855(11)	&0.00625(9)	&0.00067(5)	&0.00000	&0.00000 \\
	Co1	&0.00544(16)	&0.00402(19)	&0.00651(18)	&0.00023(13)	&0.00000	&0.00000 \\
	Co2	&0.00418(16)	&0.00320(19)	&0.00611(19)	&-0.00021(12)	&0.00000	&0.00000 \\
	Ga1	&0.0091(2)	&0.0054(2)	&0.0058(2)	&0.00188(17)	&0.00000	&0.00000 \\
	Ga2	&0.00546(15)	&0.00645(17)	&0.00790(17)	&-0.00113(11)	&0.00000	&0.00000 \\
	Ga3	&0.00424(14)	&0.00734(18)	&0.00761(17)	&0.00001(11)	&0.00000	&0.00000 \\ 
	Ga4	&0.00674(16)	&0.00413(16)	&0.01095(18)	&0.00070(12)	&0.00000	&0.00000 \\
	Ga5	&0.00488(14)	&0.00625(16)	&0.00736(15)	&0.00061(11)	&.00000	&0.00000 \\ 
	Ga6	&0.00595(15)	&0.00736(18)	&0.00714(16)	&-0.00183(12)	&0.00000	&0.00000 \\
	Ga7	&0.0051(2)	&0.0069(2)	&0.0120(2)	&-0.00142(17)	&0.00000	&0.00000 \\
	Ga8	&0.00829(15)	&0.00546(17)	&0.00737(17)	&0.00194(12)	&0.00000	&0.00000 \\
	Ga9	&0.00499(15)	&0.0093(2)	&0.01039(19)	&0.00204(12)	&0.00000	&0.00000 \\
	\hline
	\end{tabular}
\end{table}	

\begin{figure}[t]
	\includegraphics[width=0.8\textwidth]{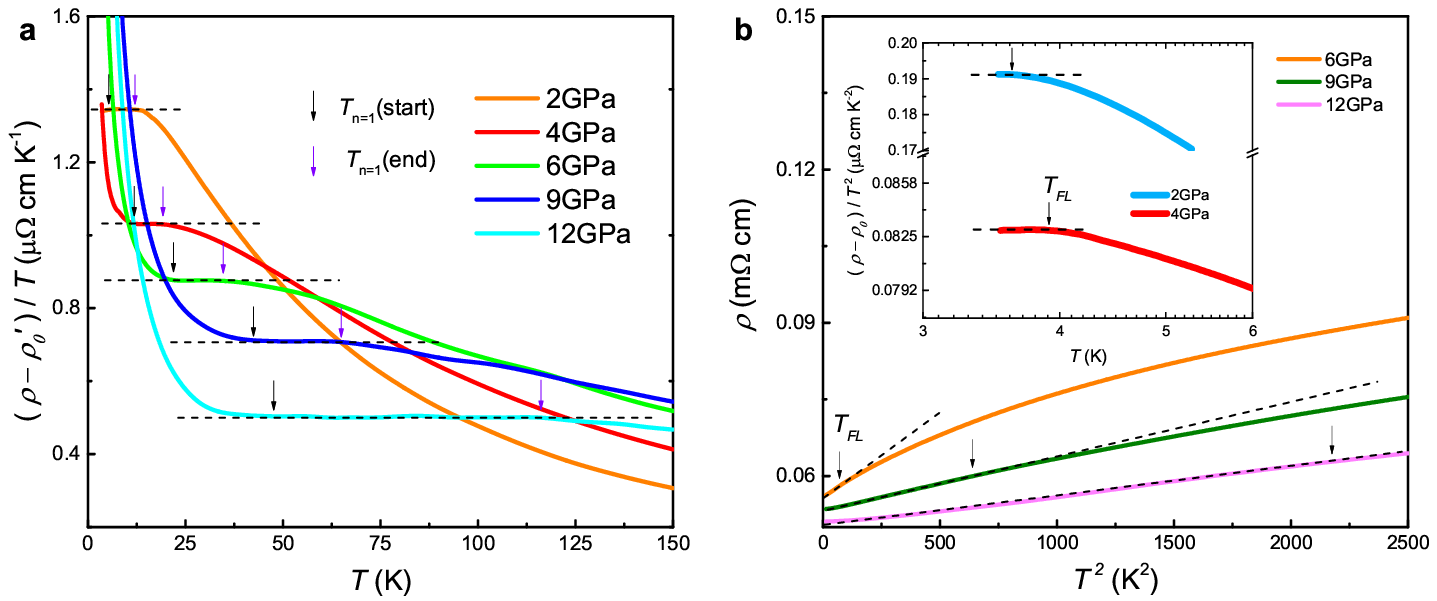}
	\caption{\label{fig1}(Color online) (\textbf{a}) The curves of $(\rho-\rho_0')/T$ versus $T$ where $\rho_0'$ is obtained from the linear fit. The plateaus suggest the $T$-linear behavior corresponding to the region for $n=1$. (\textbf{b}) The $\rho-T^2$ plot fitting with $\rho=\rho_0+A T^2$ where $\rho_0$ is the residual resistivity at zero temperature. The inset shows the temperature dependence of $(\rho-\rho_0)/T^2$ with $\rho_0$ modified properly as the residual resistivity.}
\end{figure}


\begin{thebibliography}{99}
\section*{REFERENCES}
\bibitem{White2015} White, B. D., Thompson, J. D. $\&$ Maple, M. B. Unconventional superconductivity in heavy-fermion compounds. \textit{Physica C} \textbf{514}, 246-278 (2015).
\bibitem{Steglich2016} Steglich, F. $\&$ Wirth, S. Foundations of heavy-fermion superconductivity: lattice Kondo effect and Mott physics. \textit{Rep. Prog. Phys.} \textbf{79}, 084502 (2016).
\bibitem{Yang2016} Yang, Y.-F. Two-fluid model for heavy electron physics. \textit{Rep. Prog. Phys.} \textbf{79}, 074501 (2016).
\bibitem{Coleman2005} Coleman, P. $\&$ Schofield, A. J. Quantum criticality. \textit{Nature} \textbf{433}, 226-229 (2005).
\bibitem{Si2010} Si, Q. $\&$ Steglich, F. Heavy fermions and quantum phase transitions. \textit{Science} \textbf{329}, 1161-1166 (2010).
\bibitem{Lohneysen2007} L\"ohneysen, H. V., Rosch, A., Vojta, M. $\&$ W\"olfle, P. Fermi-liquid instabilities at magnetic quantum phase transitions. \textit{Rev. Mod. Phys.} \textbf{79}, 1015 (2007). 
\bibitem{Stockert2011} Stockert, O. $\&$ Steglich, F. Unconventional quantum criticality in heavy-fermion compounds. \textit{Annu. Rev. Condens. Matter Phys.} \textbf{2}, 79-99 (2011).
\bibitem{Coleman2007} Coleman, P. Heavy fermions: electrons at the edge of magnetism, in \textit{Handbook of Magnetism and Advanced Magnetic Materials} (eds Kronmuller, H. \& Parkin, S.) Vol. {\bf 1}, 95-148 (John Wiley and Sons, 2007).
\bibitem{Schollwock2005} Schollw\"ock, U. The density-matrix renormalization group. \textit{Rev. Mod. Phys.} \textbf{77}, 259-315 (2005).
\bibitem{Stewart2001} Stewart, G. R. Non-Fermi-liquid behavior in d- and f-electron metals. \textit{Rev. Mod. Phys.} \textbf{73}, 797-855 (2001); \textbf{78}, 743-753 (2006).
\bibitem{Krellner2011} Krellner, C. \textit{et al.} Ferromagnetic quantum criticality in the quasi-one-dimensional heavy fermion metal YbNi$_4$P$_2$. \textit{New J. Phys.} \textbf{13}, 103014 (2011).
\bibitem{Steppke2013} Steppke, A. \textit{et al.} Ferromagnetic quantum critical point in the heavy-fermion metal YbNi$_4$(P$_{1-x}$As$_x$)$_2$. \textit{Science} \textbf{339}, 933-935 (2013).
\bibitem{Koterlin1989} Koterlin, M. D., Morokhivski\v\i, B. S., Lapunova, R.V. $\&$ Sichevich, O. M. New Kondo lattices of the CeM$_2$X$_8$ type (M = Fe, Co; X = Al, Ga). \textit{Sov. Phys. Solid State} \textbf{31}, 1826-1827 (1989).
\bibitem{Fert1987} Fert, A. $\&$ Levy, P. M. Theory of the Hall effect in heavy-fermion compounds. \textit{Phys. Rev. B.} \textbf{36}, 1907-1916 (1987).
\bibitem{Bonner1964} Bonner, J. C. $\&$ Fisher, M. E. Linear magnetic chains with anisotroyic coupling. \textit{Phys. Rev.} \textbf{135}, A640-A658 (1964).
\bibitem{Kondo1964} Kondo, J. Resistance minimum in dilute magnetic alloys. \textit{Prog. Theor. Phys.} \textbf{32}, 37-49 (1964).
\bibitem{Sidorov2002} Sidorov, V. A. \textit{et al.} Superconductivity and quantum criticality in CeCoIn$_5$. \textit{Phys. Rev. Lett.} \textbf{89}, 157004 (2002).
\bibitem{Steglich2014} Steglich, F. \textit{et al.} Evidence of a Kondo destroying quantum critical point in YbRh$_2$Si$_2$. \textit{J. Phys. Soc. Jpn.} \textbf{83}, 061001 (2014).
\bibitem{Park2011} Park, T. \textit{et al.} Unconventional quantum criticality in the pressure-induced heavy-fermion superconductor CeRhIn$_5$. \textit{J. Phys.: Condens. Matter} \textbf{23}, 094218 (2011).
\bibitem{Kittel1966} Kittel, C. \textit{Introduction to Solid State Physics} (Wiley, New York, 1966).
\bibitem{Yang2015} Yang, C. L. \textit{et al.} Kondo effect in the quasiskutterudite Yb$_3$Os$_4$Ge$_{13}$. \textit{Phys. Rev. B} \textbf{91}, 075120 (2015).
\bibitem{Prakash2016} Prakash, O., Thamizhavel, A. $\&$ Ramakrishnan, S. Ferromagnetic ordering of minority Ce$^{3+}$ spins in a quasi-skutterudite Ce$_3$Os$_4$Ge$_{13}$ single crystal. \textit{Phys. Rev. B} \textbf{93}, 064427 (2016).
\bibitem{Yang2008} Yang, Y.-F., Fisk, Z., Lee, H.-O., Thompson, J. $\&$ Pines, D. Scaling the Kondo lattice. \textit{Nature} \textbf{454}, 611-613 (2008).
\bibitem{Petrovic2001} Petrovic, C. \textit{et al.} Heavy-fermion superconductivity in CeCoIn$_5$ at 2.3 K. \textit{J. Phys.: Condens. Matter} \textbf{13}, L337-L342 (2001).
\bibitem{Blaha2001} Blaha, P., Schwarz, K., Madsen, G. K. H., Kvasnicka, D. $\&$ Luitz, J. (eds) \textit{WIEN2k: an augmented plane wave plus local orbital program for calculating the crystal properties} (Vienna University of Technology, 2001).
\bibitem{Perdew1996} Perdew, J. P., Burke, K. $\&$ Ernzerhof, M. Generalized gradient approximation made simple. \textit{Phys. Rev. Lett.} \textbf{77}, 3865-3868 (1996).
\bibitem{Koterlin1994} Koterlin, M. D., Morokhivski\v\i, B. S., Babich, N. G. $\&$ Zakharenko, N. I. Characteristic magnetic properties of the new Kondo lattice CeFe$_2$Al$_8$. \textit{Phys. Solid State} \textbf{36}, 632-634 (1994).
\bibitem{Kolenda2001} Kolenda, M. \textit{et al.} Low temperature neutron diffraction study of the CeFe$_2$Al$_8$ compound. \textit{J. Alloys Compd.} \textbf{327}, 21-26 (2001).
\bibitem{Ghosh2012} Ghosh, S. $\&$ Strydom, A. M. Strongly correlated electron behaviour in CeT$_2$Al$_8$ (T = Fe, Co). \textit{Acta Phys. Pol. A} \textbf{121}, 1082-1084 (2012).
\bibitem{Cheng2014} Cheng, J.-G. \textit{et al.} Integrated-fin gasket for palm cubic-anvil high pressure apparatus. \textit{Rev. Sci. Instrum.} \textbf{85}, 093907 (2014).
\end{thebibliography}
\end{document}